# Throughput Limits of IEEE 802.11 and IEEE 802.15.3

Sana Ullah[1], Yingji Zhong[1, 2], Riazul Islam[1], Ahasanun Nessa[1], and Kyung Sup Kwak[1]

[1]Graduate School of Information Technology and Telecommunications, Inha University
253 Yonghyun-dong, Nam-gu, Incheon, 402-751, Korea.
Email: sanajcs@hotmail.com, dubd96@yahoo.com,
anessa.inha@gmail.com, kskwak@inha.ac.kr.

[2] School of Information Science and Engineering
27, Shanda Nan Road, ,Shandong University,250100,Jinan,China.
Email: zhongyingji32@sdu.edu.cn.

*Abstract*-IEEE 802.11 and IEEE 802.15.3 are wireless standards originally designed for Wireless Local Area Network (WLAN) and Wireless Personal Area Network (WPAN). This paper studies MAC throughput analysis of both standards. We present a comparative analysis of both standards in terms of MAC throughput and bandwidth efficiency. Numerical results show that the performance of IEEE 802.15.3 transcends IEEE 802.11 in all cases.

## I. INTRODUCTION

Wireless LANs are becoming more sophisticated and innovative due to low cost, genuine mobility, fast implementation, customer satisfaction and high data rate communication. IEEE 802.11 is a wireless LAN standard originally designed for wireless communication. It defines the physical and MAC layer specifications for wireless LAN [1], where it supports two modes of operation. Distributed Coordination Function (DCF) doesn't have central node and all nodes compete for the channel using CSMA/CA protocol. Point Coordination Function (PCF) has a central node, which controls the network by broadcasting beacon frames periodically. DCF and PCF modes can coexist in one cell by defining interframe spacing [2]. Moreover, it has one MAC, which interacts with three PHYs. Frequency Hop Spread Spectrum (FHSS) operates in 2.4 GHz band, Direct Sequence Spread Spectrum (DSSS) operates in 2.4 GHz and Infrared. There are different versions of IEEE 802.11, which includes variations on physical layer. However, we focus on IEEE 802.11a - which uses Orthogonal Frequency Division Multiplexing (OFDM) technique to deliver up to 54Mbps in 5 GHz Industrial Scientific Medical (ISM) band, and IEEE 802.11b – which uses High Rate Direct Sequence Spread Spectrum (HR-DSSS) technique to deliver up to 11Mbps in 2.4 GHz ISM band.

In contrast to wireless LAN, wireless PAN (Personal Area Network) contains a number of independent wireless devices connected within a small range around a person or object. IEEE 802.15.3 is a set of standards developed for high data rate wireless PAN [3]. The wireless PAN network is controlled by a central coordinator called piconet, which controls the medium and maintains network synchronization timing via beacons. The channel is bounded by superframes structure given in Fig 1, where each superframe begins with a beacon and consists of three components: the beacon - which transmits control information to piconet, Contention Access Period (CAP) – which uses CSMA/CA mechanism to communicate commands or asynchronous data, and Contention Free Period (CFP) – which uses TDMA protocol, where devices are assigned specified time slots for isochronous streams. The piconet coordinator can sometime replace CAP with Management Time Slots (MTS) using slotted Aloha access scheme. It uses Quadrature Amplitude Modulation (64-QAM) technique to deliver up to 55Mbps in 2.4 GHz ISM band. A UWB physical layer for wireless PAN delivers up to 480 Mbps in 3.1-10.6 GHz band [4].

This paper presents throughput analysis of both standards provided by MAC layer, i.e., maximum number of MSDUs transmitted in a unit time. We compare the bandwidth efficiency of both standards. The rest of the paper is categorized into three sections. In Section 2, we discuss about the throughput calculation of both standards. Section 3 presents the numerical results, and finally we present conclusion to our work.

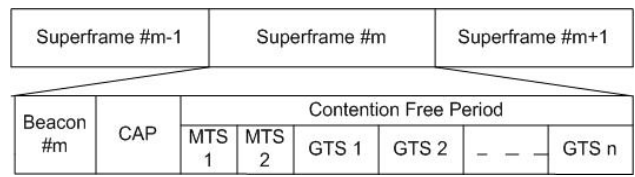

Figure 1: IEEE 802.15.3 piconet superframe

## II. THROUGHPUT CALCULATION

The maximum throughput is defined as the maximum number of MAC Layer Service Data Units (MSDUs) that are transmitted in a unit time. We consider the MAC layer throughput comparison of both standards. Each MSDU carries additional overhead at MAC and Physical layer such as PHY preambles and MAC headers, control frames, interframe spacing and back-off time in case of IEEE 802.11. In IEEE 802.11, the overhead is transmitted at control rate while in IEEE 802.15.3, overhead is transmitted at basic data rate of 22Mbps. Our calculation considers all the assumptions defined in [5], i.e., there are no collisions in case of IEEE 802.11, the transmission is error free and at least one station has always a packet to send. In the following section, we present numerical calculations to derive the maximum throughput of IEEE 802.11 and IEEE 802.15.3.

### A. MAC Throughput of IEEE 802.11

In IEEE802.11, data frames and control frames are transmitted at different rates. In case of CSMA/CA, the station transmits if the channel is free for Distributed Interframe Spacing period (DIFS). Short Interframe Spacing (SIFS) framing is used to separate transmission belong to a single dialog. Each frame in IEEE 802.11 is composed of additional delay created by interframe spacing and back off period. In case of RTS/CTS, the transmission cycle contains 3 SIFS period in addition to RTS and CTS frames.

The maximum throughput of IEEE 802.11 is presented in [6], where the upper throughput limit of IEEE 802.11a and IEEE 802.11b is derived. However, the derivation ignores RTS/CTS mechanism. The RTS/CTS mechanism is considered in [5], where the throughput is analyzed for payload size of 4000 bytes but the propagation delay is ignored. Our throughput calculation of IEEE 802.11 is based on the formulas given in [5] and [6] with a slight modification in the MAC header $H_{MAC}$ and the addition of propagation delay $\tau$ in case of RTS/CTS mechanism. All other parameters are taken from the standard [1].

The maximum throughput $M_T$ is calculated as the ratio of payload size $x$ to the transmission delay per payload size and is given by:

$$M_T = \frac{8x}{Delay} \quad (1)$$

Where

$$Delay(x) = T_{DIFS} + T_{BO} + T_{RTS} + T_{CTS} + T_{Data} + T_{SIFS} + T_{ACK} + 2\tau \quad (2)$$

The transmission time of RTS/CTS is zero in case of CSMA/CA. The data transmission time $T_{Data}$ for IEEE 802.11b is given by:

$$T_{Data} = T_{PHY} + T_P + \frac{8(H_{MAC} + x)}{R} \quad (3)$$

$T_{Data}$ for IEEE 802.11a is given by:

$$T_{Data} = T_{PHY} + T_P + T_{SYM} * ceil(\frac{(22 + 8(H_{MAC} + x))}{N_{DBPS}}) \quad (4)$$

where $H_{MAC}$ = 30bytes.

### B. MAC Throughput of IEEE 802.15.3

In IEEE 802.15.3, the channel is divided into superframes, with each superframe having three components, i.e., beacon, optional CAP and CFP. For throughput derivation, we ignore beacon and optional CAP period. We only consider the Channel Time Allocation Period (CTAP) based on TDMA access scheme. In CTAP, users are assigned specified time slots called Channel Time Allocations (CTAs) by the piconet coordinator. Three acknowledgement policies are defined in the standard [3]. Imm-ACK is issued for each data frame, Dly-ACK is issued for the burst of frames and No-ACK means no acknowledgement for the data frame. The use of Dly-ACK improves channel utilization by reducing the frequency of ACK and SIFS frames.

The data transmission time of a frame is given by:

$$T_{Data} = T_{PHY} + T_P + (T_{MAC} + T_{HCS}) + (8 \times x / R) + T_{FCS} + T_{STUFF} + T_{TAIL} \quad (5)$$

The maximum throughput for Imm-ACK is derived as

$$M_{Imm-ACK} = \frac{8x}{T_{Data} + T_{Imm-ACK} + 2 \times T_{SIFS}} \quad (6)$$

The maximum throughput for Dly-ACK is derived as

$$M_{DLY-ACK} = \frac{8x}{(nT_{Data} + T_{Imm-ACK} + (n-1)T_{MIFS} + 2 \times T_{SIFS})/n} \quad (7)$$

In IEEE 802.15.3, MAC and PHY headers are transmitted at basic data rate of 22Mbps, while payload including Frame Check Sequence (FCS), tail symbols and stuff bits are transmitted at desired data rate. Tail symbols are added to end of the MAC frame. For 11Mbps (QPSK-TCM format), 3 tail symbols are added to the end of MAC frame. For 16/32/64-QAM formats, 4 tail symbols are added to the MAC frame. If the size of MPDU (payload plus FCS) is not an integer multiple of bits/symbol, then stuff bits are added to the MAC frame. The number of stuff bits should be less than the number of bits contained in a symbol. Moreover, enough stuff bits should be added so that the MPDU plus stuff bits is an integer multiple of the bits/symbol. Stuff bits are added only to 33Mbps (3 bits/symbol) and 55Mbps (5 bits/symbol) modes. In this case, MPDU is not an integer multiple of 3 and 5 i.e.

$$\frac{8\times(Payload+FCS)}{5}=1.6\times(Payload+FCS) \qquad (8)$$

Hence, 0.4 is added as stuff bits. For 11Mbps (1 bit/symbol), 22Mbps (2 bits/symbol) and 44Mbps (4 bits/symbol) modes, there is no need of stuff bits.

### III. NUMERICAL RESULTS

The parameters for both standards are given in table 1. The numerical results show that throughput and bandwidth efficiency of IEEE 802.15.3 is higher than those of IEEE 802.11. Fig 2 shows the maximum throughput of IEEE 802.15.3 for Imm-ACK and Dly-ACK. For the payload size of 1000 bytes and 55Mbps data rate, maximum throughput is 44Mbps and 48Mbps for Imm-and Dly-ACK, respectively. The Dly-ACK is issued for the burst of 5 frames. Fig 3 presents the maximum throughput of IEEE 802.11a. For 54Mbps, the throughput is 25Mbps without RTS/CTS and 20Mbps with RTS/CTS. The use of RTS/CTS decreases the throughput performance. The Throughput Upper Limit (TUL) of IEEE 802.11a is 50.2 Mbps [6]. Due to large overhead in IEEE 802.11, the throughput doesn't not exceed TUL, even for higher data rate of 100,000 Mbps. Fig 4 shows the MAC throughput and TUL of IEEE 802.11b for various data rates. Fig 5 and 6 presents the comparative analysis of both standards in terms of throughput and bandwidth efficiency. Bandwidth efficiency is the ratio of maximum throughput $M_T$ to the desired data rate $R$ and is given by:

$$BE=\frac{M_T}{R} \qquad (9)$$

The throughput performance of IEEE 802.15.3 is higher than that of IEEE 802.11a in all cases. For 55Mbps and Imm-ACK, the throughput of IEEE 802.15.3 is 44Mbps while the throughput of IEEE 802.11a is 25Mbps for 54Mbps. The performance of IEEE 802.11 is significantly influenced by additional overhead such as back off and control frames. The bandwidth efficiency of IEEE 802.15.3 is 80% for 55Mbps while its only 46% in IEEE 802.11a.

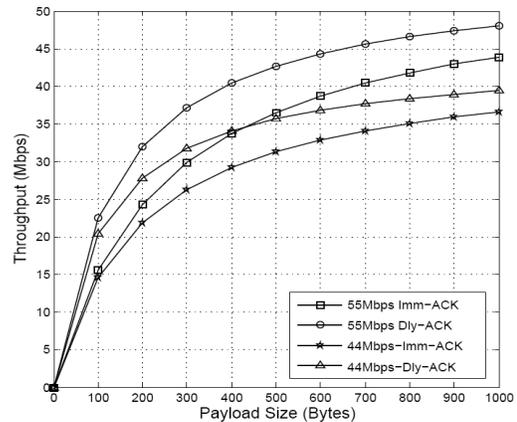

Figure 2. Maximum MAC throughput of IEEE 802.15.3

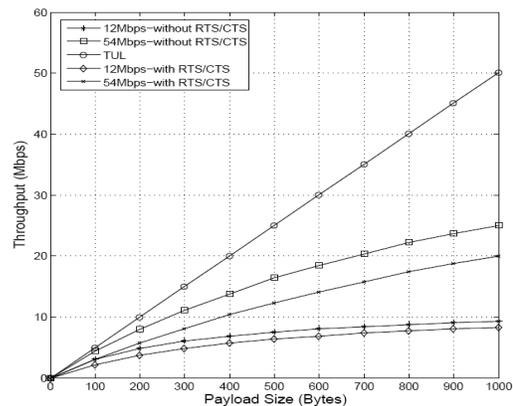

Figure 3. Maximum MAC throughput of IEEE 802.11a- with RTS/CTS

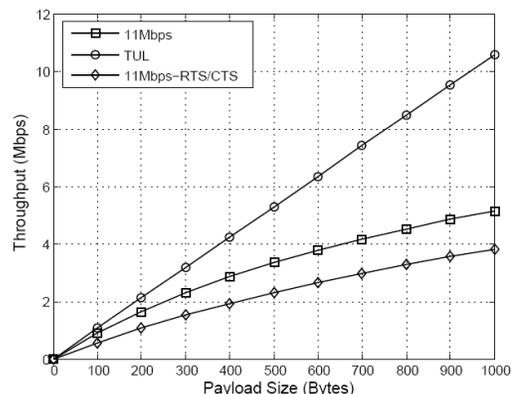

Figure 4.Maximum MAC throughput of IEEE 802.11b with RTS/CTS

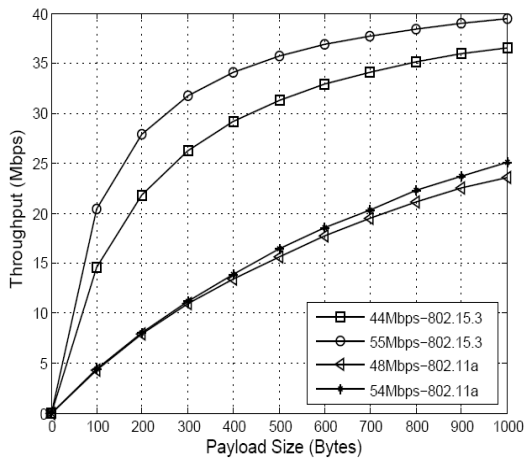

Figure 5. Throughput Comparison of IEEE 802.15.3 and IEEE 802.11a

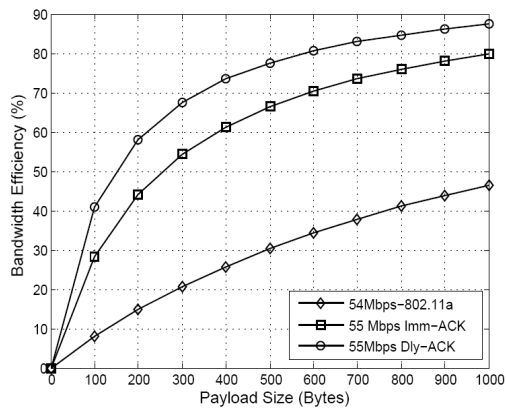

Figure 6. Bandwidth Efficiency Comparison of IEEE 802.15.3 and IEEE 802.11a

## IV. CONCLUSIONS

In this paper, we presented MAC throughput analysis of IEEE 802.11 and IEEE 802.15.3 standards. We calculated the theoretical throughput limit of both standards. Our numerical results urge the use of Dly-ACK for high traffic. The comparative analysis of both standards concluded that the performance of IEEE 802.15.3 transcends IEEE 802.11 in terms of throughput and bandwidth efficiency. However, the adaptation of any standard depends on the application and customer requirements. For instance, IEEE 802.15.3 is suitable for home theater, interactive video gaming and high speed video transfer.

In the future, the performance of both standards will be compared in terms of power consumption. The comparison of numerical and simulation results is also part of our future work.

TABLE 1: PARAMETERS OF IEEE 802.11 AND IEEE 802.15.3

|  | IEEE 802.11a | IEEE 802.11b | IEEE 802.15.3 |
|---|---|---|---|
| $T_{PHY}$ | 4 $\mu$s | 48 $\mu$s | 0.727 $\mu$s |
| $T_P$ | 16 $\mu$s | 144 $\mu$s | 7.27 $\mu$s |
| $T_{MAC}$ | 8*30/R | 30*30/R | 3.63 $\mu$s |
| $T_{HCS}$ | - | - | 0.727 $\mu$s |
| $T_{STUFF}$ | - | - | 4 bits |
| $T_{TAIL}$ | - | - | 4 bits |
| $T_{SIFS}$ | 16 $\mu$s | 10 $\mu$s | 10 $\mu$s |
| $T_{MIFS}$ | - | - | 2 $\mu$s |
| $T_{DIFS}$ | 34 $\mu$s | 50 $\mu$s | - |
| $T_{BO}$ | 15 | 31 | - |
| $\tau$ | 1 $\mu$s | 1 $\mu$s | - |
| $T_{SYM}$ | 4 $\mu$s | - | - |

ACKNOWLEDGMENTS

This research was supported by the MKE (Ministry of Knowledge Economy), Korea, under the ITRC (Information Technology Research Center) support program supervised by the IITA (Institute of Information Technology Assessment) (IITA-2008-C1090-0801-0019).